\documentclass[letter,twocolumn]{jpsj2}
\title{Re-parameterization Invariance in Fractional Flux Periodicity}

\author{Shuichi \textsc{Murakami}
\thanks{E-mail address: murakami@appi.t.u-tokyo.ac.jp},
Ken-ichi \textsc{Sasaki}$^{1}$
\thanks{E-mail address: sasaken@imr.tohoku.ac.jp} and
Riichiro \textsc{Saito}$^{2}$
\thanks{E-mail address: rsaito@flex.phys.tohoku.ac.jp}}

\inst{
Department of Applied Physics, University of Tokyo,
Hongo, Bunkyo-ku, \\
Tokyo 113-8656, Japan\\
$^{1}$ Institute for Materials Research, Tohoku University, 
Sendai 980-8577, Japan \\
$^{2}$ Department of Physics, Tohoku University and CREST, JST,\\
Sendai 980-8578, Japan
}

\abst{
 We analyze a common feature of a nontrivial fractional flux periodicity
 in two-dimensional systems.
 We demonstrate that an addition of fractional flux can be absorbed into 
 re-parameterization
 of quantum numbers. 
For an exact fractional periodicity, all the 
 electronic states undergo the re-parameterization, whereas for 
an approximate periodicity valid in a large system, 
 only the states near the Fermi level are involved in the re-parameterization.
}

\kword{Aharonov-Bohm effect, carbon nanotube, fractional flux periodicity,
persistent current, Fermi surface}

\begin{document}

\maketitle

The Aharonov-Bohm (AB) effect~\cite{AB,Tonomura} is one of the direct 
manifestations of the  quantum nature of electrons.
The interference pattern obtained in an AB experiment shows that a single
electron wave function has the fundamental unit of magnetic flux, 
$\Phi_0=hc/e$.
The AB effect has important physical consequences also in solid state
physics.
For example, an equilibrium persistent current~\cite{Buttiker,Cheung2}, 
which is
a derivative of the ground state energy $E_0(\Phi)$ by the threading flux
$\Phi$, $I_{\rm pc} = - \frac{\partial E_0(\Phi)}{\partial \Phi}$ in
mesoscopic metallic rings is observed 
experimentally~\cite{Levy,Chandrasekhar,Mailly}.
Coherence effects between electrons appear in $I_{\rm pc}$ as 
functions of threading flux.
Since the ground state of materials consists of many electrons, the AB 
effect can lead to physical results due to the coherence between the
electrons.
One of the interesting coherence effects is a fractional flux periodicity
in the ground state energy.
The fractional flux periodicity means that for $\Delta = \Phi_0/Z$ ($Z$
is an integer), $E_0(\Phi+\Delta)=E_0(\Phi)$ holds exactly or in a certain 
limit for any $\Phi$.
There are several theoretical studies on the fractional flux periodicity.
Cheung et al.~\cite{Cheung} found that a finite length cylinder with a
specific aspect ratio exhibits the fractional flux periodicity in the
persistent currents.
The same configuration obtained with a magnetic field applied perpendicular
to the cylindrical surface was shown to have  a fractional flux
periodicity by Choi and Yi~\cite{Choi}.
Such cylinders are composed of a square lattice.
In addition to these cylinders, the 
torus geometry composed of a square lattice exhibits
the fractional flux periodicity, depending on the twist around the torus
axis and the aspect ratio~\cite{SKS,SKS2}. 
Similarly to a square lattice, a honeycomb lattice can also show the
fractional flux periodicity.  
We found that an armchair carbon nanotube with heavy doping can
exhibit the fractional flux periodicity~\cite{SMS}. 
Even though all these systems showing a fractional flux periodicity are
two-dimensional (2D) systems, common features for the fractional
periodicity have not yet been clarified.

If all electronic states for $\Phi$ and $\Phi
+ \Delta$ have the same energy in one-to-one
correspondence, the AB effect can occur. However,
when we plot one electron energy as a function of
magnetic field, each electron state for $\Phi$
does not always correspond to the state for $\Phi+\Delta$ with
the same energy; the state for $\Phi+\Delta$ with
the same energy may come from the other state for
$\Phi$. In this case, we can specify the states for
$\Phi+\Delta$ by re-parameterizing the quantum numbers
of the states for $\Phi$. A general question is
whether there is such a re-parametrizing operation
as a function of magnetic field. In this context, we have
shown for some fractional periodic systems that
there is a
re-parametrizing operation that gives an
invariance (re-parametrization invariance) for
a single-electron energy. 
There are two types of fractional flux periodicity; one is 
an exact one, while the other is achieved in a limit of a large system.
For both types of fractional flux periodicity, in general, an
addition of fractional flux can be recognized as a
re-parameterization of quantum numbers, as we 
will discuss later on a twisted torus~\cite{SKS,SKS2} and a
cylinder~\cite{Cheung} composed of a square lattice.
For the exact periodicity, all the electronic states for $\Phi$ and 
$\Phi+\Delta$ are in one-to-one correspondence by the 
re-parameterization, whereas for the approximate periodicity, only the 
states near the Fermi level are involved in the re-parameterization.

We first analyze the flux periodicity of a twisted torus composed of a
square lattice considered in refs.~\citen{SKS,SKS2}.
We consider a nearest-neighbor tight-binding model on a torus, with 
the hopping integral $t$ between nearest-neighbor sites.
Let $N$ and $Q$ denote the numbers of lattice sites around and along the
torus axis, respectively, and 
let $\delta N$ denote the twist along the torus axis.
The energy eigenvalue of the system is 
\begin{eqnarray}
&& E_{\mu_1 \mu_2}(\Phi) = 
 - 2 t \left\{
 \cos \left( \frac{2\pi \mu_1}{N} \right)\right.
\nonumber \\
&&\ \ \ \ 
 +\left. \cos\frac{2\pi}{Q}\left(\mu_2- \frac{\delta
 N}{N}\mu_{1}-\frac{\Phi}{\Phi_0}\right)
 \right\}, 
 \label{eq:t-torus}
\end{eqnarray}
where $\mu_1$ and $\mu_2$ are integer quantum numbers,
taking the values $\mu_{1}=1,\cdots,N$, $\mu_{2}=1,\cdots Q$.
Because $E_{\mu_1,\mu_2}=E_{\mu_1+N,\mu_{2}+\delta N}=E_{\mu_1,\mu_{2}+Q}$,
the region for the integers $\mu_{1}$ and $\mu_{2}$ can be 
taken as any $N$ and $Q$ consecutive integers, respectively.
When the Fermi level is fixed at $E_F =0$, $E_{0}(\Phi)$ can be expressed 
as $
 E_0(\Phi) = \sum'_{\mu_1 \mu_2} E_{\mu_1 \mu_2}(\Phi)$,
where $\sum'$ is a summation over the states with negative energy
eigenvalues $E_{\mu_1 \mu_2}(\Phi)<0$.
We can rewrite it as
\begin{eqnarray}
E_{0}(\Phi)&=&
 -\frac{1}{2} \sum_{\mu_1 \mu_2} 
 \left( \left| E_{\mu_1 \mu_2}(\Phi)
 \right|-E_{\mu_1 \mu_2}(\Phi) \right)\nonumber \\
&=&-t \sum_{\mu_1, \mu_2} 
 \left| \cos \left( \frac{2\pi \mu_1}{N} \right)\right. \nonumber \\
&&\ \ \ \ \ \left.
 + \cos \frac{2\pi}{Q} \left(\mu_2 - \frac{\delta
 N}{N}\mu_1-\frac{\Phi}{\Phi_0}\right) \right|,
 \label{eq:t-g-ene}
\end{eqnarray}
because eq.~(\ref{eq:t-torus}) implies 
$\sum_{\mu_1 \mu_2}E_{\mu_1 \mu_2}=0$ (electron-hole symmetry).
Hence, if $\Delta$ is a flux periodicity of the ground state energy, 
eq.~(\ref{eq:t-g-ene}) yields
\begin{eqnarray}
 &&\sum_{\mu_1, \mu_2}\left|\cos\frac{2\pi\mu_{1}}{N}
  +\cos\frac{2\pi}{Q}\left(\mu_{2}-
 \frac{\delta N}{N}\mu_{1}-\frac{\Phi}{\Phi_{0}}\right)\right|= \nonumber \\
 &&\sum_{\mu'_1, \mu'_2}\left|\cos\frac{2\pi\mu'_{1}}{N}
 +\cos\frac{2\pi}{Q}\left(\mu'_{2}-
 \frac{\delta N}{N}\mu'_{1}-\frac{\Phi+\Delta}{\Phi_{0}}\right)\right|
\nonumber \\
&&
 \label{eq:identity}
\end{eqnarray}
for an arbitrary $\Phi$.
Let us check that $\Phi_0$ is an exact period of the system~\cite{BY},
i.e., eq.~(\ref{eq:identity}) is satisfied for $\Delta=\Phi_{0}$.
By setting $\mu_1' =
\mu_1$ and $\mu_2' = \mu_2 + 1$, the summands of
both sides of eq.~(\ref{eq:identity}) become equal,
which is the re-parameterization operation of the system.
Moreover, 
since the region of $\mu_2$ can be taken as any $Q$ 
consecutive integers as noted previously, 
this shift of $\mu_{2}$ does not 
affect the result.
Therefore, the ground state energy has a $\Phi_0$ periodicity and 
the translation of $\mu_{2}$ gives the
re-parameterization invariance of $E_{0}(\Phi)$.

Now we examine if the system has another flux periodicity in addition to 
that of $\Phi_0$.
For eq.~(\ref{eq:identity}) to hold for an arbitrary $\Phi$, 
the summations on both sides should be termwise equal.
Hence, there should be one-to-one correspondence between $(\mu_1,\mu_2)$
and $(\mu'_1,\mu'_2)$, 
and either of 
the following two conditions should hold:
\begin{eqnarray}
 &\text{(i)}&\ 
 \cos\left(\frac{2\pi\mu_{1}}{N} \right)=
 \cos\left(\frac{2\pi\mu'_{1}}{N} \right), \\
 && \ \cos\frac{2\pi}{Q}\left(\mu_{2}- \frac{\delta
 N}{N}\mu_{1}-\frac{\Phi}{\Phi_{0}}\right)
\nonumber \\
&& \ \ \ \  =\cos\frac{2\pi}{Q}\left(\mu'_{2}- \frac{\delta
 N}{N}\mu'_{1}-\frac{\Phi+\Delta}{\Phi_{0}}\right), 
\end{eqnarray}
or 
\begin{eqnarray}
 &\text{(ii)}&\ 
 \cos\left(\frac{2\pi\mu_{1}}{N} \right)=
 -\cos\left(\frac{2\pi\mu'_{1}}{N} \right),\\ 
 &&\cos\frac{2\pi}{Q}\left(\mu_{2}-\frac{\delta
 N}{N}\mu_{1}-\frac{\Phi}{\Phi_{0}}\right) 
\nonumber \\
&& \ \ \ \  =-\cos\frac{2\pi}{Q}\left(\mu'_{2}-\frac{\delta
 N}{N}\mu'_{1}-\frac{\Phi+\Delta}{\Phi_{0}}\right). 
\end{eqnarray}
The case (i) leads to 
$\mu_1' \equiv \mu_1 \ ({\rm mod}\ N)$. Let us take $\mu'_{1}=\mu_{1}$, 
resulting in 
$\mu_2' \equiv \mu_2 + \Delta/\Phi_0 \ ({\rm mod}\ Q)$. 
This condition is satisfied when $\Delta$ is an integer multiple of $\Phi_{0}$.
This corresponds to the normal AB effect for this system with 
a $\Phi_{0}$ periodicity.

On the other hand, the case (ii) can lead to the nontrivial periodicity of
$E_0(\Phi)$. 
This in turn leads to $\mu'_1 \equiv \mu_{1} + N/2 \  ({\rm mod}\ N)$,
which 
is allowed for even $N$. Let us suppose $N$ is even, and we obtain
$\mu'_{1}=\mu_{1}+N/2$
and $\mu'_2 \equiv \mu_{2}+(Q+\delta N)/2+ \Delta/\Phi_0 \ ({\rm mod}\
Q)$. 
There are two distinct cases; 
(ii-a) if $Q+\delta N$ is even, only an integer multiple of $\Phi_0$ is
allowed for $\Delta$, and (ii-b)
if $Q+\delta N$ is odd, $\Phi_0/2 $ is
allowed for $\Delta$.
The case (ii-a) leads to the trivial $\Phi_{0}$ periodicity, while
the case (ii-b) leads to the nontrivial $\Phi_{0}/2$ periodicity.
To summarize, when $N$ is even and $Q+\delta N$ is odd, 
the period is $\Phi_0/2$, and the period is $\Phi_{0}$
otherwise, in agreement with numerical results in refs.~\citen{SKS,SKS2} 
\cite{note}.

The above analysis is for the exact periodicity of $E_0(\Phi)$.
On the other hand, as pointed out in refs.~\citen{SKS,SKS2}, 
there can also be an approximate periodicity of $E_0(\Phi)$ that is less 
than the exact periodicity calculated above.
For the approximate periodicity, the 
re-parameterization transforms
the states near the Fermi level for $\Phi$ to those
for $\Phi+\Delta$.
Here we show that if $\delta N/N=p/q$ for coprime integers $p$ and $q$, 
and $Q/N$ is an integer, the period is 
$\Phi_{0}/q$, which becomes asymptotically valid for large $N$, $\delta N$
and $Q$. This result agrees with the numerical results in 
refs.~\citen{SKS,SKS2}.
To show this $\Phi_{0}/q$ periodicity, 
we should expand $E_{0}(\Phi)$ in terms of $1/Q$ and 
extract the lowest-order term dependent on $\Phi$. This 
procedure eventually corresponds to extracting 
the contribution from the electronic states near 
the Fermi level, and it is called a 
regularization procedure in a general context. 
For a regularization procedure in 
one-dimensional relativistic models, see, for example, 
refs.~\citen{Manton,Iso}.
This regularization procedure requires the linearization of 
the energy spectrum near the Fermi level and the introduction of 
energy cutoff far from the Fermi level.
While it is applicable to the present case, we can also 
derive the result directly 
without the introduction of an artificial cutoff, as
we explain briefly here.
In the present case with large $Q$, 
we first express the summation over $\mu_{2}$
by an integral 
with correction 
terms, using the formula
\begin{eqnarray}
&&\frac{1}{Q}\sum_{\mu_{2}=1}^{Q}g\left(\frac{\mu_{2}}{Q}\right)
= \int_{0}^{1}g(x)dx\nonumber \\
&&\makebox[1cm]{} +\frac{1}{12Q^{3}}\sum_{\mu_{2}=1}^{Q}
g''\left(\frac{\mu_{2}}{Q}\right)+O\left(\frac{1}{Q^{3}}\right),
\label{eq:sum-g}
\end{eqnarray}
which holds for the arbitrary differentiable function $g(x)$ with 
$g(x)=g(x+1)$.
In order to calculate eq.~(\ref{eq:t-g-ene}),
it is tempting to substitute $|E_{\mu_{1}\mu_{2}}|$ for $g(\frac{\mu_{2}}{Q})$ 
in eq.~(\ref{eq:sum-g}).
The first two terms in the r.h.s.\ of eq.~(\ref{eq:sum-g}) then turn 
out be independent of $\Phi$, and they 
do not contribute to the persistent current. 
In fact, however, we have left out another contribution.
The summand $|E_{\mu_{1}\mu_{2}}|$ is not differentiable with respect to
$x=\mu_{2}/Q$ when $E_{\mu_{1},\mu_{2}}=0$, 
i.e., at the Fermi level. This gives a 
finite correction to the result. 
This procedure is visualized in Fig.\ 1. 
This correction to the order $1/Q^{2}$ is 
the lowest-order term dependent on $\Phi$, namely,
it gives the leading-order term for the persistent current.
It is evaluated as
\begin{eqnarray}
&& E_0(\Phi) \approx \sum_{\mu_{1}} \frac{2\pi v_F(\mu_{1})}{Qa} 
 \frac{ Q_L(\mu_{1},\Phi) ^2 +  Q_R(\mu_{1},\Phi)
 ^2}{2}\nonumber \\
&& \ \ \ \ \ \  + \text{const.},
\label{eq:E0-regularized} 
\end{eqnarray}
where $a$ is the lattice constant and 
\begin{eqnarray}
&& Q_L(\mu_1,\Phi)\nonumber \\
&&\ \ 
= \frac{\Phi}{\Phi_{0}} -A^L(\mu_{1}) - \left[ \frac{\Phi}{\Phi_{0}} -A^L(\mu_{1}) \right] - \frac{1}{2}, \label{eq:QL}\\
&&  Q_R(\mu_1,\Phi)\nonumber \\
&&\ \  
=\left[ \frac{\Phi}{\Phi_{0}} + A^R(\mu_{1}) \right] - \left( \frac{\Phi}{\Phi_{0}} + A^R(\mu_{1}) \right)
 +\frac{1}{2}.\ \ \  \ \label{eq:QR}
\end{eqnarray}
Here, we introduce 
$A^L(\mu_{1}) \equiv 
\frac{Q}{2}+ \left(\frac{-\delta N-Q}{N} \right)\mu_{1}$, 
$A^R(\mu_{1}) \equiv \frac{Q}{2}+ \left(\frac{\delta N-Q}{N} \right)\mu_{1}$
and the
Fermi velocity $v_F(\mu_{1}) \equiv 2ta \sin \left(\frac{2\pi\mu_{1}}{N}
\right)$. 
$ Q_L + Q_R $ and $
Q_L - Q_R $ correspond to the regularized
charge and current of the $\mu_{1}$-th mode, respectively. 
When $Q/N$ is an integer and  $\delta N/N=p/q$, we obtain
\begin{eqnarray}
&&Q_L(\mu_1,\Phi) =
 Q_L\left(\mu_1-1,\Phi+\frac{p}{q}\Phi_{0}\right) 
, \label{eq:QL+1}\\
&&
 Q_R(\mu_1,\Phi) = Q_R
\left(\mu_1-1,\Phi+\frac{p}{q}\Phi_{0}\right) 
,
 \label{eq:QR+1}
\end{eqnarray}
while $v_{F}(\mu_{1}\pm 1)=v_{F}(\mu_{1})(1+O(\frac{1}{N}))$.
Hence, it follows that 
\begin{equation}
E_{0}(\Phi)=E_{0}\left(\Phi+\frac{p}{q}\Phi_{0}\right)\left(1+O\left(
\frac{1}{N}\right)\right).
\end{equation}
Thus, in the $N\rightarrow \infty$ limit, the system has
a $(p/q)\Phi_{0}$ fractional periodicity. This correspondence 
between $E_0(\Phi)$ and $E_0(\Phi+\frac{p}{q}\Phi_{0})$ is related to the
re-parameterization of the  $\mu_{1}$ 
quantum numbers (eqs.~(\ref{eq:QL+1}) and (\ref{eq:QR+1})), 
restricted to those at the Fermi level
in the present case.
Thus, we have shown the approximate $(p/q)\Phi_{0}$ flux periodicity
in the twisted torus.
Combined with the trivial $\Phi_{0}$ periodicity,
this yields a $\Phi_{0}/q$ periodicity in this case.
This is because for mutually coprime integers $q$ and $p$,
there exist integers $\alpha$ and $\beta$ such that $\alpha p+\beta q=1$,
yielding $\alpha(p/q)\Phi_{0}+\beta\Phi_{0}=\Phi_{0}/q$.
\begin{figure}[htb]
\includegraphics[scale=0.65]{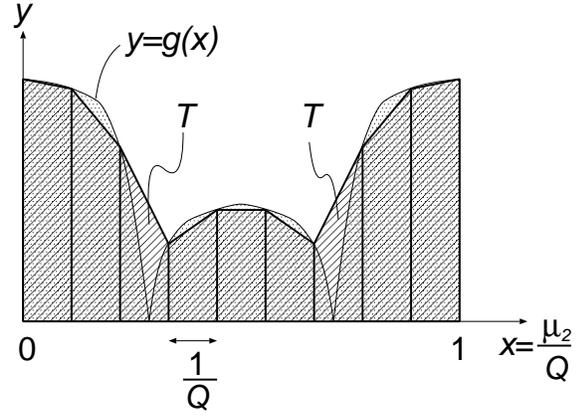}
\caption{
Schematic picture for an expansion of $\sum_{\mu_2}|E_{\mu_{1}\mu_{2}}(\Phi)|$
in terms of $1/Q$.
This procedure is expressed as eq.~(\ref{eq:sum-g}), and 
is regarded as an approximation of the curve 
$y=g(\frac{\mu_{2}}{Q})\equiv|E_{\mu_1,\mu_2}|$ by 
a collection of segments.
Each term in eq.\ (\ref{eq:sum-g})
can be associated with an area of a certain region in the figure.
The hatched and  
dotted regions represent the l.h.s.\ 
and the first term of the r.h.s.\ of eq.~(\ref{eq:sum-g}), 
respectively. Their difference to the order $1/Q^{2}$ (the second term
of the r.h.s.\ of eq.~(\ref{eq:sum-g})) is represented 
by narrow arcs between the curve $y=g(x)$ and the segments.
There is an additional 
contribution near the points with $E_{\mu_1 \mu_2}=0$, i.e.,\ 
from the Fermi level.
This comes from the triangular regions, shown as ``$T$'' in the figure, 
and results in the $\Phi$-dependent term shown in 
eq.\ (\ref{eq:E0-regularized}) 
to the order $1/Q^{2}$.}
\end{figure}

Next, we consider a two-dimensional cylinder composed of a square
lattice~\cite{Cheung,SMS}.
We again consider a nearest-neighbor tight-binding model with 
the hopping integral $t$.
This model can also exhibit the fractional flux
periodicity \cite{Cheung,SMS}.
Let $N$ ($M$) denote the number of lattice sites along (around) the
cylindrical axis.
The cylinder does not have the twist degree of freedom.
The energy eigenvalue of the system is 
\begin{equation}
 E_{n_1 n_2}(\Phi) = -2t 
 \left\{ 
 \cos \left( \frac{n_1 \pi}{N+1} \right) 
 + \cos \frac{2\pi}{M}\left( n_2 - \frac{\Phi}{\Phi_0} \right)
 \right\},
 \label{eq:cylin-ene}
\end{equation}
where $n_1$ and $n_2$ are integer quantum numbers and $1 \le n_1 \le N$ and
$1\leq n_{2}\leq M$.
An exact fractional periodicity $\Delta$ of the ground state energy 
imposes the following equation for any $\Phi$:
\begin{eqnarray}
 && \sum_{n_1, n_2} \left| \cos \left( \frac{n_1 \pi}{N+1} \right) 
 + \cos \frac{2\pi}{M}\left( n_2 - \frac{\Phi}{\Phi_0} \right) \right|
 \nonumber \\
 &&\ =  \sum_{n'_1, n'_2} \left| \cos \left( \frac{n'_1 \pi}{N+1} \right) 
 + \cos \frac{2\pi}{M}\left( n'_2 - \frac{\Phi+\Delta}{\Phi_0}
 \right)\right|.\nonumber \\
&&
 \label{eq:cylin-all}
\end{eqnarray}
This requires a transformation between $(n'_1,n'_2)$ and
$(n_1,n_2)$, which can absorb the fractional flux $\Delta$ in
eq.~(\ref{eq:cylin-all}). 
By an analysis similar to that of the twisted torus, we conclude that,
for a nontrivial flux periodicity, we should use the re-parameterization
$n'_{1}=N+1-n_{1}$ and $n'_{2}= n_{2}+\frac{M}{2}+\frac{\Delta}{\Phi_{0}}$,
which yields a $\frac{\Phi_{0}}{2}$ periodicity for odd $M$.

For an approximate periodicity, 
only the states 
near the Fermi level are relevant to the flux-dependent 
part of the ground state energy to the order $1/M$.
In the limits $N\rightarrow\infty$ and 
$M\rightarrow \infty$ with the fixed integer $Z=2(N+1)/M $, 
$E_0(\Phi)$ has a fractional flux periodicity of $\Delta =
\Phi_0/Z$~\cite{Cheung,SMS} . 
The ground state energy to the order $1/M$ is given by
\begin{equation}
 E_0(\Phi) \approx \sum_{n_1} \frac{2\pi v_F(n_1)}{Ma} 
 \frac{ Q_L(n_1,\Phi) ^2 +  Q_R(n_1,\Phi)
 ^2}{2} + \text{const.}, 
\end{equation}
where $ Q_{L}$ and $ Q_{R}$ 
are defined by  eqs.~(\ref{eq:QL}) and (\ref{eq:QR}) with 
$A^{L}(n_1)= A^{R}(n_1) 
\equiv \frac{M}{2} \left(1-\frac{n_1}{N+1} \right)$ and the
Fermi velocity $v_F(n_1) \equiv 2ta \sin \left(\frac{n_1 \pi}{N+1}
\right)$. 
When $2(N+1)/M=Z$ is an integer, we can see that
the system has a $\Phi_0 /Z$ periodicity in the $N\rightarrow \infty $ 
limit. This is similar to that in the case of the  
twisted torus, and this corresponds to the
re-parameterization of quantum numbers of states at the Fermi level:
$n'_1 = n_1+ 1$ for $Q_{L}$ and $n'_1 = n_1 -1$ for $Q_{R}$.

On the basis of the results of the above analyses of the two systems, 
we can have
further insight into the general aspects of fractional flux periodicity.
For the exact fractional periodicity $\Delta$, the 
electron-hole symmetry plays 
a crucial role; if one breaks the symmetry by shifting the 
Fermi energy from zero or by allowing a next-nearest-neighbor hopping,
the exact fractional periodicity disappears, and only the trivial 
$\Phi_{0}$ periodicity remains.
On the other hand, the approximate periodicity in the limit of a large 
system is more robust, since it involves only the states near the
Fermi level. 
The approximate fractional flux periodicity is determined from 
$A^{L}$ and $A^{R}$, which reflects the 
Fermi wavenumbers. This might be a key to understand experimental 
results on an approximate fractional flux periodicity in the 
magnetoresistance
of carbon nanotubes \cite{Bachtold,Fujiwara}, for which 
some theoretical studies \cite{Roche,Roche2}
have
been carried out. A complete explanation of the experimental
results is under way.

In summary, we have shown 
that the fractional flux periodicity is a result of the
re-parameterization invariance. 
This means that
if the system has the fractional flux
periodicity $\Delta$, the additional fractional flux $\Delta$ 
can be absorbed by a
translation of quantum numbers.
For the exact periodicity, 
it transforms all the states.
Meanwhile, for the 
approximate periodicity, asymptotically valid in large systems, 
the re-parameterization involves  only the states 
at the Fermi level.

 S. M. is supported by a Grant-in-Aid (No.~16740167) from the Ministry of
 Education, Culture, Sports, Science and Technology, Japan.
K. S. is supported by a fellowship of the 21st Century COE Program of
the International Center of Research and Education for Materials of
 Tohoku University. 
 R. S. acknowledges Grants-in-Aid (Nos.~13440091 and 16076201) 
from the Ministry of
 Education, Culture, Sports, Science and Technology, Japan.


\begin{thebibliography}{99}
\bibitem{note}
It should be mentioned that, for some lattice structures, the Fermi level
$E_F=0$ is not compatible with the conservation of the number of
states below the Fermi level when we vary the AB flux.
Let us give a simple example. In a one-dimensional tight-binding model with
$N$ lattice sites, an energy eigenvalue is given by $E_{\mu_1}
= -2t \cos \frac{2\pi}{N}(\mu_1+N_\Phi)$.
When we fix the Fermi level as $E_F=0$, the number of states
below the Fermi level is given by $Q \equiv \left[ N/4 - N_\Phi \right]
- \left[ -N/4 - N_\Phi \right]$.
Therefore, for odd $N$, $Q$ is not a constant.
We can assume that the chemical potential of the system is fixed at
$E_F=0$ to make the system open and applicable to
transport experiments.
\end{thebibliography}
\end{document}